\documentclass[amssymb,amsmath,aps,prl,twocolumn]{revtex4}
\usepackage{graphicx}
\usepackage{dcolumn}
\usepackage{bm}
\usepackage[up]{subfigure}

\begin{document}


\title{Dipole trap for $^{87}$Rb atoms using lasers of different wavelength}


\author{C. M. \surname{Chandrashekar}}%

\affiliation{Atomic and Laser Physics,
             University of Oxford,
             Oxford OX1 3PU, UK}

\begin{abstract}

The parity of atomic wave functions prevents neutral atoms from having 
permanent electric-dipole moment. Electric-dipole moment is induced in an 
atom when exposed to strong light, the electric field of the light.
Hence the optical trapping of neutral atoms relies on the induced dipole 
moment. Here we present the calculated numerical values of the detuning, potential depth, minimum laser power required to trap $^{87}$Rb ($D_2$ line) atoms using lasers of wavelength 1064 nm, 850 nm, 820 nm and 800 nm and beam waists 50$\mu$m, 100$\mu$m and 200$\mu$m.

\end{abstract}

\maketitle

\section{Introduction}

Trapping and storage of charged and neutral particles have laid pathways
for advancement in physics~\cite{chu}. Neutral atoms are trapped on the basis 
of magnetic and optical interactions. Magnetic traps represent ideal conservative traps and arise from the state-dependent force on the permanent magnetic dipole moments in inhomogeneous field, restricting the experiments to only few special case. Optical dipole traps~\cite{grimm}, which is much weaker than other trapping mechanism rely on the electric dipole interaction with far-detuned light. Under appropriate conditions, the dipole trapping mechanism is independent of the particular sub-level of the electronic ground-state. The internal ground state can thus be fully exploited for experiments which is not possible in magnetic trapping mechanism.\\

\section{Frequency-detuning and potential depth for $^{87}Rb$ atoms} 

Effective laser detuning $\Delta$ for alkali's can be calculated using,
\begin{equation}
\frac{1}{\Delta}=\left(\frac{1}{\Delta_{1}}+\frac{2}{\Delta_{2}}\right)
\end{equation}
\noindent where $ \Delta_{i} $ is detuning from $D_{i}$ line.\\

For an alkali-metal atom, the maximum potential depth is calculated 
from~\cite{Cho},
\begin{equation}
U_{0}=\frac{\hbar\Gamma}{2}\frac{P\Gamma}{\pi W_{0}^{2}I_{0}\Delta}
\end{equation}
\noindent where $\Gamma=1/\tau$, is the natural linewidth and
$I_{0}$, the saturation intensity, given by,
\begin{equation}
I_{0}=\frac{\pi^{2}h c
\Gamma}{3\lambda^{3}}=\frac{\pi}{3}\frac{hc}{\lambda^{3}\tau}.
\end{equation}

The numerical value of saturation intensity $I_{0}$ for $D_{2}$ line
($|F=2, m_{F} = \pm2\rangle \rightarrow |F^{\prime}=3,
m_{F^{\prime}}=\pm 3 \rangle $) atomic transition of $^{87}$Rb
atoms is calculated to be 1.67 mW/cm$^{2}$=16.7 W/m$^{2}$ .

\section{Frequency-detuning and potential depth using lasers of different wavelength}

Below is the table with numerical values of detuning and potential
depth for dipole traps using laser lights of different
wavelength.\\

\begin{tabular}{|c|c|c|}
\hline
wavelength & Detuning & Potential Depth $U_{0}$ \\
$nm$       &  $\Delta$ &    $W/m^{2}$ \\
\hline
$1064$ & $-25\times10^{5}\Gamma$ & $1.53\times10^{-35}\frac{P}{W_{0}^{2}}$\\
$850$ & $-6.93\times10^{5}\Gamma$ & $5.5\times10^{-35}\frac{P}{W_{0}^{2}}$\\
$820$ & $-3.46\times10^{5}\Gamma$ & $11.04\times10^{-35}\frac{P}{W_{0}^{2}}$\\
$800$ & $-0.6\times10^{5}\Gamma$ & $63.66\times10^{-35}\frac{P}{W_{0}^{2}}$\\
\hline
\end{tabular}\\

Atoms in an intense laser field experience an ac start shift. This shift 
creates a potential $U$ proportional to the light intensity, such that

\begin{equation}
U(r,z)=U_{0}\left[\frac{e^{\frac{-2r^{2}}{W(z)^{2}}}}
{1+(\frac{z}{z_{R}})^{2}}\right]
\end{equation}

\noindent at the focus of a Gaussian laser beam, with $r$ and $z$ being the radial and axial co-ordinates, $z_{R}=\frac{\pi W_{0}^{2}}{\lambda}$ is the Rayleigh
range \footnote{Distance at which the diameter of the spot size
increases by a factor of $\sqrt{2}$} at wavelength $\lambda$ and
beam waist $W_{0}$. $W(z)$ is the beam radius as a function of
axial position $z$ and is given by,

\begin{equation}
W(z)=W_{0}\sqrt{1+\left(\frac{z}{z_{R}}\right)^{2}}.
\end{equation}

\noindent For a beam in horizontal direction the overall potential
due to effect of gravity is given by,

\begin{equation}
U_{g}(r,z)= - mgr -
U_{0}\left[\frac{e^{\frac{-2r^{2}}{W(z)^{2}}}}{\left(1 +
\frac{z}{z_{R}}\right)^{2}}\right]
\label{pot}
\end{equation}

\section{Minimum power required to trap $^{87}Rb$ atoms at distance $z$ from
the focal point of the beam}

To calculate the minimum power required to trap $^{87}$Rb atoms at
distance $z$ from the focal point of the beam in the axial
direction, the above equation (Eq.~\ref{pot}) is differentiated,

\begin{equation}
\frac{d^{2}U}{dr^{2}}= -mg+\frac{4U_{0}r}{W(z)^{2}}\left[\frac{e^
{\frac{-2r^{2}}{W(z)^{2}}}}{1 + (\frac{z}{z_{R}})^{2}}\right]=0
\end{equation}

\begin{equation}
\Longrightarrow r =
\frac{mg}{4U_{0}}\frac{W(z)^{2}}{e^{\frac{-2r^{2}}{W(z)^{2}}}}
\left[1+\left(\frac{z} {z_{R}}\right)^{2}\right] \label{diffpot}
\end{equation}

By substituting the appropriate values for the above equation
(Eq.~\ref{diffpot}) one can calculate the minimum power required
to trap atoms at distance $z$ from the beam focus in the axial
direction. Below is the table with the calculated power required
to trap $^{87}$Rb atoms using laser light with different
wavelengths and beam waist, 50 $\mu m$,
100 $\mu m$ and 200 $\mu m$.\\

For light with beam waist, $W_{0}=50 \mu$m \\

\begin{tabular}{|c|c|c|c|}
\hline
Distance from focus & 1064 nm & 850 nm & 820 nm\\
\hline
$z=$ 5 cm & 3.33 W & 450 mW & 210 mW\\
$z=$ 2 cm & 30 mW & 37 mW & 26.5 mW\\
$z=$1 cm & 16 mW & 8.4 mW & 4 mW\\
$z=$ 0.5 cm & 11.6 mW & 3.8 mW & 1.5 mW\\
$z=$ 0 cm & 9 mW  & 2.7 mW & 1.4 mW\\
\hline
\end{tabular}\\
\\
\\
for $W_{0}=100 \mu$m \\

\begin{tabular}{|c|c|c|c|}
\hline
Distance from focus & 1064 nm & 850 nm & 820 nm\\
\hline
$z=$ 5cm & 610 mW & 110 mW & 47 mW\\
$z=$ 2cm & 140 mW & 31 mW & 15 mW\\
$z=$ 1cm & 91 mW & 23 mW & 12 mW\\
$z=$ 0.5cm & 78 mW & 22 mW & 11 mW\\
$z=$ 0cm & 77 mW  & 22 mW & 11 mW\\
\hline
\end{tabular}\\
\\
\\
for $W_{0}=200 \mu$m \\

\begin{tabular}{|c|c|c|c|}
\hline
Distance from focus & 1064nm & 850nm & 820nm\\
\hline
$z=$ 5cm & 1.53 W & 200 mW & 1.1 W\\
$z=$ 2cm & 650 mW & 180 mW & 900 mW\\
$z=$ 1cm & 610 mW & 180 mW & 900 mW\\
$z=$ 0.5cm & 600 mW & 180 mW & 900 mW\\
$z=$ 0cm & 600 mW  & 180 mW & 900 mW\\
\hline
\end{tabular}\\

The maximum photon scattering rate $\Gamma_{sc}$ is given by
\begin{equation}
\Gamma_{sc}=\frac{\Gamma}{\Delta}\frac{U_{0}}{\hbar}
\end{equation}

\end{document}